\documentclass{article}
\usepackage{spconf,amsmath,graphicx}

\usepackage{url}
\usepackage{mathrsfs}
\usepackage{fourier}

\usepackage{rotating}
\usepackage{bbm}
\usepackage{latexsym}
\newtheorem{definition}{Definition}

\title{Robust Dependence Measure using RKHS based Uncertainty Moments and Optimal Transport}
\name{Rishabh Singh and Jose C. Principe\thanks{This work was partially supported by DARPA under grant no. FA9453-18-1-0039 and ONR under grant no. N00014-21-1-2345.}}
\address{University of Florida}
\begin{document}
%
\maketitle
\begin{abstract}
Reliable measurement of dependence between variables is essential in many applications of statistics and machine learning. Current approaches for dependence estimation, especially density-based approaches, lack in precision, robustness and/or interpretability (in terms of the type of dependence being estimated). We propose a two-step approach for dependence quantification between random variables: 1) We first decompose the probability density functions (PDF) of the variables involved in terms of multiple local moments of uncertainty that systematically and precisely identify the different regions of the PDF (with special emphasis on the tail-regions). 2) We then compute an optimal transport map to measure the geometric similarity between the corresponding sets of decomposed local uncertainty moments of the variables. Dependence is then determined by the degree of one-to-one correspondence between the respective uncertainty moments of the variables in the optimal transport map. We utilize a recently introduced Gaussian reproducing kernel Hilbert space (RKHS) based framework for multi-moment uncertainty decomposition of the variables. Being based on the Gaussian RKHS, our approach is robust towards outliers and monotone transformations of data, while the multiple moments of uncertainty provide high resolution and interpretability of the type of dependence being quantified. We support these claims through some preliminary results using simulated data.
\end{abstract}
\section{Introduction}
\label{sec:intro}

Accurate and domain-specific dependence measures between time-series has always been an important research problem with many applications, notably in finance related areas. Conventionally used correlation measures (Pearson correlation for instance) have many disadvantages such as inability to effectively capture non-linear dependencies and lack of robustness to noise and monotone transformations \cite{mart1}. They also lack in their interpretability of the type of dependence being quantified \cite{sur}. This has, to some extent, been addressed by ``equitable'' measures of dependence (which treat linear and non-linear dependence in the same manner) where the intrinsic idea is to measure the distance $D[P(X,Y), P(X)P(Y)]$ between joint distribution $P(X, Y)$ and product of marginals $P(X)P(Y)$ of variables $X$ and $Y$ \cite{chang, reshef, szek}. Mutual information and copulas are popular examples of such measures \cite{mi, cop, mart1}. Copulas particularly are known for efficiently disentangling the shape of marginals from the effects due to dependence between variables involved. However, they only capture the strength of dependence and have many practical difficulties in high dimensional problems \cite{sur}. Moreover, mutual information and copula based approaches are inadequate in separating/classifying the types of dependence, which is critical in many application domains. For instance, when evaluating dependencies in stock market data, tail-dependence is given much higher priority than others since one is more interested in knowing whether two stock market variables are correlated at their respective extreme values than at their means \cite{fin, matt, mart2}. Hence there is a need to come up with dependence measures that are both robust-equitable (i.e. able to efficiently quantify non-linear dependence while being robust to noise/transformations), and also decomposable (i.e. able to cluster different types of dependencies specific to a problem domain) for a better intuitive understanding of the types of dependence involved between variables.

\begin{figure}[!t]
    \centering\includegraphics[scale = 0.15]{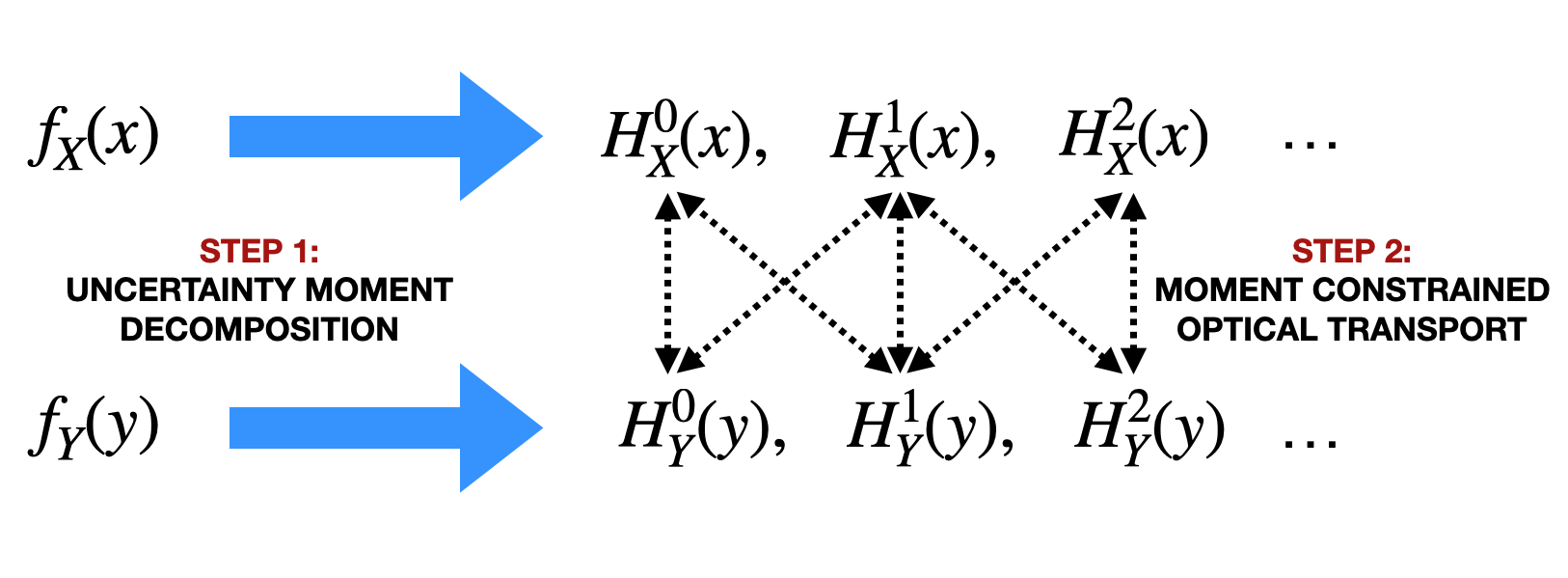}
    \caption{Approach for quantification of dependence between two random variables ($X$ and $Y$): (1). The QIPF framework is used for extracting uncertainty moments corresponding to variable marginal PDFs. (2). Optimal Transport is implemented to obtain coupling map between uncertainty moments of the variables}
    \label{apr}
\end{figure}

To this end, we propose a new decomposable and robust-equitable measure for dependence between variables involving a two-step approach. First, we deconstruct the marginal probability density functions (PDFs) of variables involved, $f_X(x)$ and $f_Y(y)$, in terms of sets of moments $\big\{H_X^0(x), H_X^1(x), H_X^2(x), ...\big\} \in H_X$ and $\big\{H_Y^0(y), H_Y^1(y),\\ H_Y^2(y), ...\big\} \in H_Y$ respectively so that each of these moments signifies a different region of the variable PDF based on the degree of uncertainty at that region (with successively higher order moments signifying the more uncertain regions of the variable in sample space). For this, we use a recently introduced framework for uncertainty decompositon called the Quantum Information Potential field (QIPF) that has been shown to be very effective for model uncertainty quantification \cite{me1, me2} and well as for time-series data uncertainty decomposition \cite{me3}. In the next step, we use moment-constrained optimal transport \cite{ot} to obtain a transportation coupling map between the two uncertainty moment sets $H_X$ and $H_Y$. Dependence is then quantified by measuring how much the coupling map deviates from one-to-one correspondence of the moment sets. This approach is depicted in Fig. \ref{apr}. For perfect correlation between $X$ and $Y$, we expect the coupling map between the uncertainty moment sets to be only concentrated at the diagonals, i.e. each moment from one marginal transports to the same corresponding mode in the other marginal (indicating heterogeneous correspondence between marginals). As correlation strength decreases, the one-to-one correspondence of the mode sets should also decrease starting from the higher order modes, which respectively reveal the more heterogeneous (extreme valued) tail-regions of the marginals. Hence correspondence between higher order moments indicate tail-dependence between the variables. Such a framework will therefore allow us to choose the type of dependence (relevant to a particular problem) based on which uncertainty moment we decide to analyze. The Gaussian RKHS (on which the QIPF is based) further makes the overall dependence measure robust-equitable because of its well known properties of representing non-linear relationships linearly (in the kernel space) while being robust to outliers and noise \cite{smola, emb}.

\section{QIPF Uncertainty Decomposition Framework}
\label{sec:unc}
In this section, we describe the QIPF framework for uncertainty decomposition of data samples associated with a random variable and also refer the reader to \cite{me2, me1} where the framework derivation and intuition is described in detail for the application of model uncertainty quantification.

The QIPF decomposition framework for data uncertainty quantification relies on measuring the variability of data PDF around a particular test sample. The intuition here is that a lack of samples corresponding to the local space around the test point (and therefore greater uncertainty) will lead to high variability of the data PDF around it. The goal is therefore to quantify the \textit{local gradient flow} (or heterogeneity) of $f_X(x)$ around the test sample $x$. To this end, the first step is to use an RKHS based function called the \textit{information potential field} \cite{prin} to estimate $f_X(x)$. It is defined as follows:

\begin{definition}[Information Potential Field]
Given a non-empty set $X = \Omega \subset \mathbb{R}^d$ representing an input random variable and a symmetric non-negative definite kernel function $K: X \times X \rightarrow \mathbb{R}$, the information potential field (IPF) is an estimator of the PDF of $X$ in the RKHS from $n$ samples ($x_1, x_2, ..., x_n$) where $x_i \in X$ as

  \begin{equation}
 \psi_{X}(x) = \frac{1}{n}\sum_{t=1}^{n}K(x_t, x).
 \label{ipff}
  \end{equation}
  

\end{definition}

At its core, the IPF is the RKHS equivalent of the Parzen's window method \cite{parz} which is a non-parametric estimator of a continuous density function $f_X(x)$ in an asymptotically unbiased form directly from data (i.e. satisfying the condition that $\lim_{n\to\infty} E[f_X(x_n)] = f_X(x)$, where $n$ is the number of samples). We choose $K$ as the Gaussian kernel in the IPF formulation without loss of generality towards other symmetric non-negative definite functions and henceforth we will denote the kernel as $G(.,.)$. The IPF is therefore an efficient RKHS based asymptotically unbiased estimator of the input data PDF defined empirically by samples. In the RKHS, it is a functional that exists over the projected sample space, in the form of $\psi_X(.)$ which gets estimated at any point $x$ in the input space through kernel evaluations becoming the scalar $\psi_X(x)$. 


\begin{figure}[!t]
    \centering\includegraphics[scale = 0.3]{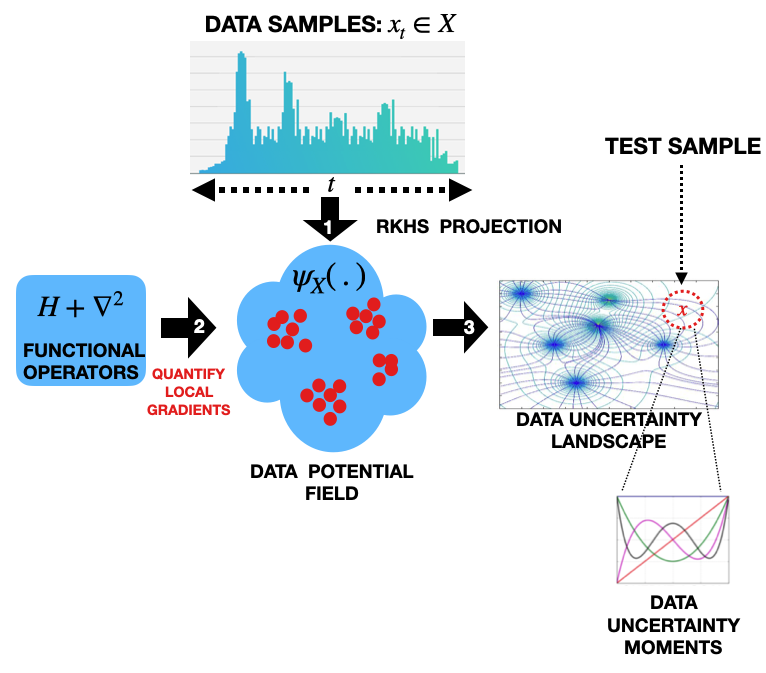}
    \caption{QIPF Framework for Uncertainty Decomposition.}
    \label{qipf}
\end{figure}

The proposed methodology calls for a more specific information, the variability of $\psi_X(x)$ around $x$, which unfortunately is not readily quantified in probability theory. A useful operator for this goal is the Laplacian operator which is a spatial high-pass filter and formally measures the flow of the gradient in a potential field \cite{huang}. Our next step is therefore to utilize the Laplacian operator to measure the local gradient flow of $\psi_X(x)$ across the sample space. Using a Schr\"odinger's equation (from quantum mathematics) with a Laplacian based wave-function, we obtain an elegantly normalized and scaled form of $\nabla_x^2\psi_X(x)$, i.e. local gradient flow of $\psi_X(x)$, which we call the Quantum Information Potential Field (QIPF). It is defined as follows:

\begin{definition}[Quantum Information Potential Field]
Given a non-empty set $X = \Omega \subset \mathbb{R}^d$ representing an input random variable so that its PDF evaluated at any point $x$ is given by $\psi_X(x) = \frac{1}{n}\sum_{t=1}^{n}G_\sigma(x_t, x)$, where $n$ is the number of samples, $G$ denotes Gaussian kernel with its kernel width $\sigma$, we define a Laplacian based Schr\"odinger equation associated with $\psi_X(x)$ as follows:

\begin{equation}
    \bigg(H_X(x) - \frac{\sigma^2}{2}\nabla_{x}^2\bigg)\psi_X(x) = E_X(x)\psi_X(x)
    \label{sch2}
\end{equation}

where $H_X(x)$ is the potential energy component of the Hamiltonian and $-\frac{\sigma^2}{2}\nabla_x^2$ is the gradient flow operator component of the Hamiltonian consisting of the Laplacian operator $\nabla_x^2$ which acts on $\psi_X(x)$ with respect to sample space $x$. Upon rearranging the middle and last terms, we obtain a normalized expression for the potential energy function at any test-point $x$ given by:

\begin{equation}
    H_X(x) = E_X(x) + (\sigma^2/2)\frac{\nabla_x^2\psi_X(x)}{\psi_X(x)}
\end{equation}

which is called the quantum information potential field (QIPF). Here, $E_X(x) = -min(\sigma^2/2)\frac{\nabla_x^2\psi_X(x)}{\psi_X(x)}$ and acts as a lower bound to ensure $H_X(x)$ is always positive.

\end{definition}

The QIPF is a therefore a functional operator in the RKHS that takes the normalized Laplacian of the IPF to describe the local dynamics (rate of change) of $\psi_X(x)$ and any point $x$. It has a small value in parts of the RKHS that have high density of samples and increases its value in parts of the space that are sparse in projected data. The QIPF operator is the RKHS version of the energy based learning methods proposed by LeCun \cite{lecun2006}, which  avoids the problems associated with estimating the normalization constant in probabilistic models. A close analysis of the mathematical form of the QIPF operator shows its similarity with the integrand of Fisher's information \cite{flego}. For uncertainty quantification, we are much more interested in the low density regions of the PDF. Hence, just like in quantum mechanics, a local decomposition of the QIPF will be very beneficial for uncertainty quantification. Our final step therefore to obtain a solution (or decompose) (\ref{sch2}) in a way that induces orthogonal moment expansion of the potential function $H_X(x)$, given by $H_X^k(x)$, corresponding to moments of the wave-function $\psi_X^k(x)$ and total energy at each moment $E_X^k(x)$ where $k = 0, 1, 2, ...$ is the moment number. $H_X^k(x)$, $\psi_X^k(x)$ and $E_X^k(x)$ preserve the same meaning as $H_X(x)$, $\psi_X(x)$ and $E_X(x)$ respectively, but for different corresponding moments as denoted by $k$, each of which specializes in a different part of data PDF potential field. Formally, this decomposition of $H_X$ is defined as follows:

\begin{definition}[QIPF Decomposition]
Given a non-empty set $X = \Omega \subset \mathbb{R}^d$ representing an input random variable so that its PDF evaluated at any test point $x$ is given by $\psi_X(x) = \frac{1}{n}\sum_{t=1}^{n}G_\sigma(x_t, x)$, where $n$ is the number of samples, $G$ denotes Gaussian kernel with its kernel width $\sigma$, the uncertainty associated with the particular test sample $x$ can be described by an ordered set of $m$ decomposed weight-QIPF moments $\{H_X^1(x), H_X^2(x), ..., H_X^m(x)\}$, the $k^{th}$ moment of which is given by the expression:

\begin{equation}
    H_X^k(x) = E_X^k(x) + (\sigma^2/2)\frac{\nabla_x^2\psi_X^k(x)}{\psi_X^k(x)}
    \label{final}
\end{equation}
where, \\
$\psi_X^k(x)$: $k^{th}$ order Hermite function (normalized) projection of the information potential field $\psi_X(x)$. \\
$\nabla_x^2$: Laplacian operator acting with respect to test point $x$.\\
$E_X^k(x) = -min(\sigma^2/2)\frac{\nabla_x^2\psi_X^k(x)}{\psi_X^k(x)}$: lower bound to ensure each moment is positive.
\end{definition}

$H_X^1(x)$, $H_X^2(x)$, $H_X^3(x)$ ... are thus the successive local moments of model uncertainty (called QIPF moments) that induce anisotropy in the PDF space and represent functional measurements at $x$ corresponding to \textit{different degrees of heterogeneity of data PDF} at $x$. Successively higher order moments of the QIPF signify the more uncertain regions of the PDF where its gradient flow is high. Definitions 1-3 therefore fully describe the QIPF uncertainty decomposition framework for a random variable represented by a discrete set of samples. The framework is also depicted in Fig. \ref{qipf}.\par

\begin{figure*}[!t]
    \centering\includegraphics[scale = 0.3]{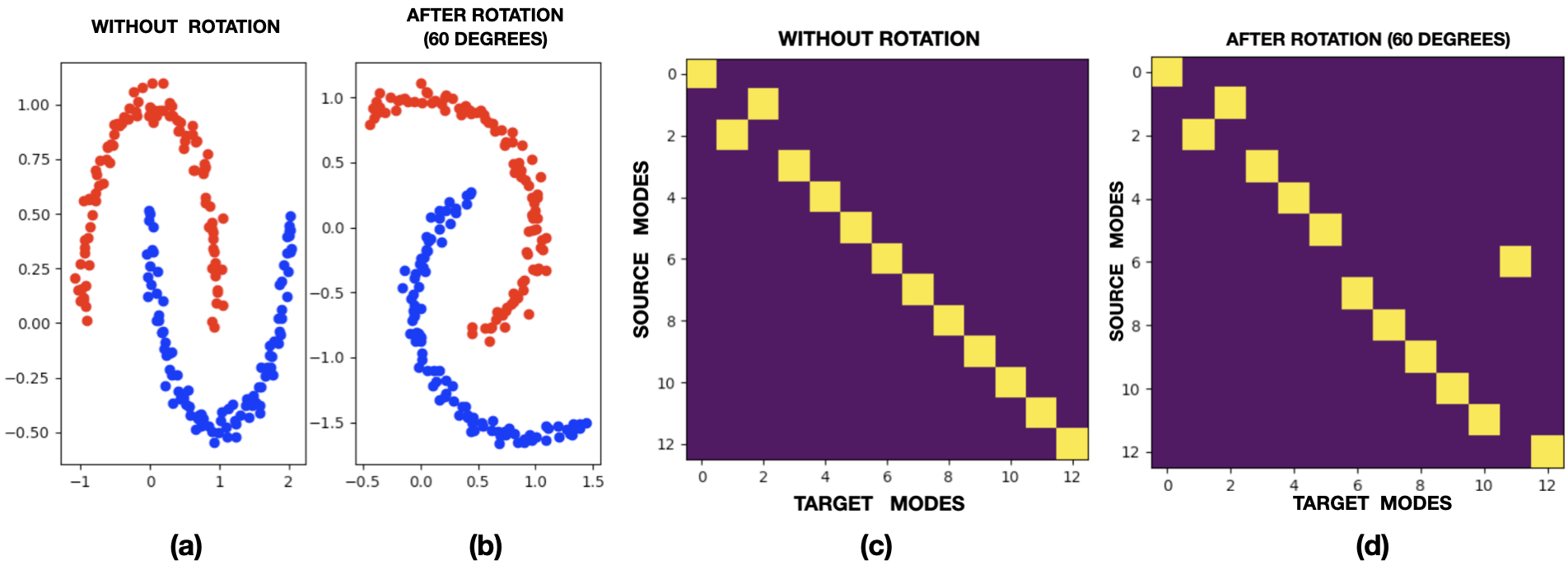}
    \caption{Robustness of QIPF-OT towards rotation transformation.}
    \label{1}
\end{figure*}

\begin{figure*}[!t]
    \centering\includegraphics[scale = 0.32]{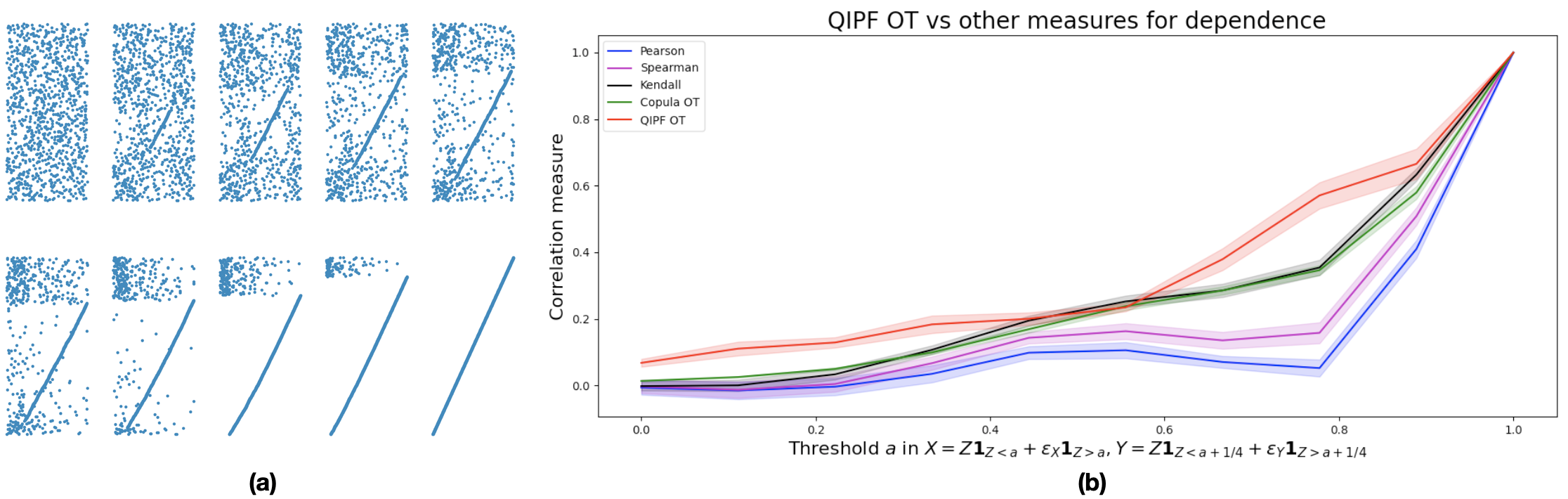}
    \caption{Consistent behavior (equitability) of QIPF-OT with different degrees of dependence as compared to other measures.}
    \label{2}
\end{figure*}

\section{Optimal Transport Implementation}
\label{sec:ot}
Our next step is to evaluate the moment-constrained optimal transport (OT) coupling matrix which is computed by taking the source and target classes as the collection of QIPF moments, $\{H_X^1(x), H_X^2(x), ..., H_X^m(x)\}$ and $\{H_Y^1(y), H_Y^2(y), ..\\., H_Y^m(y)\}$ respectively. We call it moment-constrained because we solve the OT minimization problem by adding a class (moment) based regularization term that penalizes couplings that match source samples of different moments to same target moments. Additionally, we also add a Laplacian regularization term that helps preserves local data structure. We refer the reader to \cite{ot} for details on optimal transport formulaiton and implementation.

The idea is to utilize the QIPF framework for identifying and clustering regions in the variable PDFs that have similar characteristics (in terms of PDF local gradient flow) and then use optimal transport to estimate the geometric distance between the corresponding clusters or coupling groups. The approach is slightly related to that used in \cite{mart1} where authors use empirical copulas and optimal transport to cluster financial time series based on their dependencies and show very effective results.

Dependency is then quantified by adding the indices of source variable moments which do not transport to the corresponding moments of target variable (i.e. they are non-diagonal elements in the OT coupling matrix).

\section{Experiments and Results}
\label{sec:res}

We show two preliminary results, one that indicates robustness of QIPF-OT to monotone transformation (using the two-moon data) and other that demonstrates the superior equitability of the framework when compared to other measures. Fig. \ref{1} shows the OT coupling matrix between the sets of 12 decomposed QIPF modes/moments of each moon in the two-moon dataset, before and after a rotation transformation was applied on the dataset. It can be seen that OT map remains robust to such a transformation with the lower order modes still remaining at the diagonal of the OT coupling matrix. There is only a very slight deviation of the higher order modes away from the diagonal (which is expected since the higher order modes are more sensitive). Fig. \ref{2} shows the results comparing the equitability of different dependence measures (including the more recent copula-OT \cite{mart1}) between $X$ and $Y$ using a numerical experiment (similar to that used in \cite{mart1}) where $X = Z\textbf{1}_{Z<a} + \epsilon_X\textbf{1}_{Z>a}$ and $Y = Z\textbf{1}_{Z<a+0.25} + \epsilon_Y\textbf{1}_{Z>a+0.25}$. Here $Z$ is uniform on [0, 1] and $\epsilon_X$ and $\epsilon_Y$ are independent noises. The dependence between $X$ and $Y$ is varied using the parameter $a$ ($a = 0$ for independence and $a=1$ for perfect dependence). The empirical copulas for increasing values of $a$ are shown in Fig. \ref{2}a. The dependence measured between $X$ and $Y$ with respect to $a$ is shown in Fig. \ref{2}b for different methods. The simulation was run 10 times and average with standard deviation envelope of each method was plotted. It is evident here that the QIPF-OT dependence measure is much more equitable than other methods by showing a near constant rate of change as $a$ is increased and also not decreasing at $a = 0.8$ unlike most other methods.




\section{Conclusion}
\label{sec:clc}
We proposed a new dependence measure between two random variables that leverages a Gaussian RKHS based uncertainty decomposition framework (the QIPF) and the concept of optimal transport. The QIPF decomposition framework enables one to access precise and specific parts of the random variable PDF (especially in the tail-regions) which when combined with optimal transport enables one to estimate very specific types of dependency measures that are more interpretable and equitable, as demonstrated from preliminary simulation experiments. Further, the use of the Gaussian RKHS makes the measure robust towards outliers and monotone transformations of data thus making it a promising metric for dependency measurements. Ee plan to carry out extensive experiments to assess the proposed framework.


\bibliographystyle{IEEEbib}
\bibliography{main}

\begin{thebibliography}{10}

\bibitem{mart1}
Gautier Marti, S{\'e}bastien Andler, Frank Nielsen, and Philippe Donnat,
\newblock ``Exploring and measuring non-linear correlations: Copulas,
  lightspeed transportation and clustering,''
\newblock in {\em NIPS 2016 Time Series Workshop}. PMLR, 2017, pp. 59--69.

\bibitem{sur}
Dag Tj{\o}stheim, H{\aa}kon Otneim, and B{\aa}rd St{\o}ve,
\newblock ``Statistical dependence: Beyond pearson’s $\rho$,''
\newblock {\em Statistical Science}, vol. 37, no. 1, pp. 90--109, 2022.

\bibitem{chang}
Yale Chang, Yi~Li, Adam Ding, and Jennifer Dy,
\newblock ``A robust-equitable copula dependence measure for feature
  selection,''
\newblock in {\em Artificial Intelligence and Statistics}. PMLR, 2016, pp.
  84--92.

\bibitem{reshef}
David~N Reshef, Yakir~A Reshef, Hilary~K Finucane, Sharon~R Grossman, Gilean
  McVean, Peter~J Turnbaugh, Eric~S Lander, Michael Mitzenmacher, and Pardis~C
  Sabeti,
\newblock ``Detecting novel associations in large data sets,''
\newblock {\em science}, vol. 334, no. 6062, pp. 1518--1524, 2011.

\bibitem{szek}
G{\'a}bor~J Sz{\'e}kely and Maria~L Rizzo,
\newblock ``Brownian distance covariance,''
\newblock {\em The annals of applied statistics}, vol. 3, no. 4, pp.
  1236--1265, 2009.

\bibitem{mi}
Justin~B Kinney and Gurinder~S Atwal,
\newblock ``Equitability, mutual information, and the maximal information
  coefficient,''
\newblock {\em Proceedings of the National Academy of Sciences}, vol. 111, no.
  9, pp. 3354--3359, 2014.

\bibitem{cop}
Paul Embrechts, Filip Lindskog, and Alexander McNeil,
\newblock ``Modelling dependence with copulas,''
\newblock {\em Rapport technique, D{\'e}partement de math{\'e}matiques,
  Institut F{\'e}d{\'e}ral de Technologie de Zurich, Zurich}, vol. 14, 2001.

\bibitem{fin}
Stephen Taylor,
\newblock ``Clustering financial return distributions using the fisher
  information metric,''
\newblock {\em Entropy}, vol. 21, no. 2, pp. 110, 2019.

\bibitem{matt}
David~S Matteson, Nicholas~A James, and William~B Nicholson,
\newblock ``Statistical measures of dependence for financial data,''
\newblock {\em Financial Signal Processing and Machine Learning}, p. 162, 2016.

\bibitem{mart2}
Gautier Marti, Frank Nielsen, Philippe Donnat, and S{\'e}bastien Andler,
\newblock ``On clustering financial time series: a need for distances between
  dependent random variables,''
\newblock in {\em Computational Information Geometry}, pp. 149--174. Springer,
  2017.

\bibitem{me1}
Rishabh Singh and Jose~C Principe,
\newblock ``Toward a kernel-based uncertainty decomposition framework for data
  and models,''
\newblock {\em Neural Computation}, vol. 33, no. 5, pp. 1164--1198, 2021.

\bibitem{me2}
Rishabh Singh and Jose~C. Principe,
\newblock ``A physics inspired functional operator for model uncertainty
  quantification in the rkhs,''
\newblock {\em arXiv preprint arXiv:2109.10888}, 2021.

\bibitem{me3}
Rishabh Singh and Jose Principe,
\newblock ``Time series analysis using a kernel based multi-modal uncertainty
  decomposition framework,''
\newblock in {\em Conference on Uncertainty in Artificial Intelligence}. PMLR,
  2020, pp. 1368--1377.

\bibitem{ot}
Gabriel Peyr{\'e}, Marco Cuturi, et~al.,
\newblock ``Computational optimal transport: With applications to data
  science,''
\newblock {\em Foundations and Trends{\textregistered} in Machine Learning},
  vol. 11, no. 5-6, pp. 355--607, 2019.

\bibitem{smola}
Alex~J Smola and Bernhard Sch{\"o}lkopf,
\newblock {\em Learning with kernels}, vol.~4,
\newblock Citeseer, 1998.

\bibitem{emb}
Krikamol Muandet, Kenji Fukumizu, Bharath Sriperumbudur, Bernhard
  Sch{\"o}lkopf, et~al.,
\newblock ``Kernel mean embedding of distributions: A review and beyond,''
\newblock {\em Foundations and Trends{\textregistered} in Machine Learning},
  vol. 10, no. 1-2, pp. 1--141, 2017.

\bibitem{prin}
Jose~C Principe,
\newblock {\em Information theoretic learning: Renyi's entropy and kernel
  perspectives},
\newblock Springer Science \& Business Media, 2010.

\bibitem{parz}
Emanuel Parzen,
\newblock ``On estimation of a probability density function and mode,''
\newblock {\em The annals of mathematical statistics}, vol. 33, no. 3, pp.
  1065--1076, 1962.

\bibitem{huang}
Kerson Huang,
\newblock {\em Statistical mechanics},
\newblock John Wiley \& Sons, 2008.

\bibitem{lecun2006}
Yann LeCun, Sumit Chopra, Raia Hadsell, M~Ranzato, and F~Huang,
\newblock ``A tutorial on energy-based learning,''
\newblock {\em Predicting structured data}, vol. 1, no. 0, 2006.

\bibitem{flego}
Silvana Flego, Felipe Olivares, Angelo Plastino, and Montserrat Casas,
\newblock ``Extreme fisher information, non-equilibrium thermodynamics and
  reciprocity relations,''
\newblock {\em Entropy}, vol. 13, no. 1, pp. 184--194, 2011.

\end{thebibliography}

\end{document}